\documentstyle[12pt,fleqn,epsf]{article}
\newcommand{\ssp}{\renewcommand{\baselinestretch}{0.9}
\large\normalsize}

\ssp
\def\ie{\hbox{\it i.e. }}        
        
\def\etal{\hbox{\it et al.}}

\def\abs#1{\left| #1\right|}

\def\ztautau{\ \hspace*{-1.4mm} \raisebox{-1.2mm}{$_{_Z}$} _{
\tau\tau}\ }
\def\nuww{\ \hspace*{-1mm} _{\nu} \raisebox{.7mm}{$_{_{WW}}$
} \ }
\def\nuws{\ \hspace*{-1mm}  _\nu \raisebox{.6mm}{$_{_W}$} _{
\sigma}\ }
\def\nusw{\ \hspace*{-1mm}  _{\nu\sigma} \raisebox{.6mm}{$_{
_W}$} \ }
\def\tfz{\ \hspace*{-1mm} _{\tau\Phi} \raisebox{.7mm}{$_{
_Z}$} \ }
\def\tzf{\ \hspace*{-1mm} _{\tau} \raisebox{.7mm}{$_{_Z}$
} _\Phi \ }
\def\wnunu{\ \hspace*{-1mm}  \raisebox{-.3mm}{$_{_W}$} _{
\nu\nu} \ }

\def\lra{\longrightarrow}
\def\beq{\begin{equation}}
\def\eeq{\end{equation}}
\def\bea{\begin{eqnarray}}
\def\eea{\end{eqnarray}}
\def\dsp#1{\displaystyle{#1}}
\def\np#1#2#3{    {\it Nucl. Phys. }{\bf #1} (19#2) #3}
\def\pl#1#2#3{    {\it Phys. Lett. }{\bf #1} (19#2) #3}
\def\pr#1#2#3{    {\it Phys. Rev. }{\bf #1} (19#2) #3}

\def\prl#1#2#3{    {\it Phys. Rev. Lett. }{\bf #1} (19#2) #3}

\def\zp#1#2#3{    {\it Zeit. f\"ur Physik }{\bf #1} (19#2) #3}

\def\mxfigura#1#2#3#4{
  \begin{figure}[hbtp]
    \begin{center}
      \epsfxsize=#1
      \leavevmode
      \epsffile{#2}
     \end{center}
    \caption{#3}
    \label{#4}
  \end{figure} }
\def\prepnuma{FTUV/95-15}
\def\prepnumb{IFIC/95--15}
\def\prepnumc{}
\def\titol{WEAK  DIPOLE MOMENTS AT ${\bf e^+e^-}$
COLLIDERS\footnote{
Lecture presented at the Ringberg Workshop: Perspectives for
electroweak interactions in $e^+e^-$ collisions,
February 5--8, 1995.}}
\def\autora { J. Bernab{\'e}u, G.A. Gonz{\'a}lez--Sprinberg
and J. Vidal}
\def\adressaa{Departament de F{\'{\i }}sica Te{\`o}rica,
Universitat de Val{\`e}ncia, \\ and IFIC, Centre Mixt Univ.
Valencia--CSIC\\ E--46100 Burjassot (Val\`encia), Spain}
\def\resum{
The weak dipole moments of leptons and quarks, \ie those
related to their $Z$--coupling, are reviewed.
Standard Model predictions   and  experimental
results may result in a stringent test for both their pointlike
structure and also for the Standard Model. Special attention is
devoted to the anomalous weak--magnetic dipole moment and to
the $CP$--violating weak--electric dipole moment.}

\def\today{\ifcase\month\or
 January\or February\or March\or April\or May\or June\or July
 \or August\or September\or October\or November\or December\fi
 \space\number\year}
\def\firstpage{\begin{titlepage}
\baselineskip 0.50cm \null
\vspace*{-1cm}
 \hfill \prepnuma\\
\null \hfill \prepnumb\\
\null \hfill \prepnumc\\
\baselineskip 0.75cm
\vskip 3.cm
\begin{center}
{\Large \bf \titol}
\vskip 2.cm
\baselineskip 0.6cm
{\bf \autora}
\vskip 1.cm
{\it \adressaa}
\vskip 2.cm
{\sc abstract}
\end{center}
\baselineskip 0.60cm
\begin{quotation} \resum
 \end{quotation}

\end{titlepage}\baselineskip 0.75cm}

\begin{document}
\firstpage
\setcounter{page}{2}

\def\tendeix#1#2#3{
      \stackrel{#2\rightarrow #3}{\longrightarrow} #1}
%-------------------------------------------------------
%
%--------------------BODY----------------
%
%--------------------------------------------------------
  \section{Introduction} The dipole moments of the
      electron and muon provide very precise tests of
      quantum electrodynamics.  The prediction obtained for
      the first time by Schwinger \cite{sch} was one of the
      most spectacular achievements of quantum field
      theories.  With the advent of LEP, an enormous variety
      of measurements lead to the confirmation of the
      Standard Model.  Nowadays, even pure weak--quantum
      corrections are been tested at one loop.
      The Standard Model predictions
      for the cross sections, widths and various asymmetries
      have been successfully confronted with measurements.

In this review article we will concentrate on tests and
predictions,
coming from the Standard Model, on the weak--dipole moments:
weak--magnetic and weak--electric ones.
Both weak--magnetic and weak--electric effective lagrangean terms
have the same
chirality flipping structure as mass terms. They receive
contributions from
electroweak radiative corrections in the Standard Model, but
also new physics
contributions may show up in them. In particular, they may provide
insight into the
origin of mass.

Dipole moments are quantum
corrections to the tree level
matrix elements. Within the Standard Model, whereas
 the anomalous weak--magnetic dipole
moment (WMDM) receives its leading value from one loop
corrections, the case for the weak--electric (WEDM) is rather
different.  Being a $CP$--violating property, it receives a
non--vanishing value through the Kobayashi--Maskawa mechanism,
and thereby only at very high order in the coupling
constant.

To study these dipole moments, two different approaches are
possible.
First of all, one can compute the Standard Model predictions
for them.
The other approach is the effective lagrangean approach.
 In this case
the dipole moments arise as low energy contributions
 from a high
energy physics scale.
The WEDM will be sensitive to new physics, whereas the
WMDM will receive contributions from electroweak radiative
corrections too. The corresponding operators in the effective
lagrangean are dimension--5 operators with the normalization to
$g/(2m_f)$; Marciano \cite{mar} has arg\"ued the very
general result that a fermion of mass $m_f$, generated at
 $\Lambda$--scale, has an
 anomalous moment WDM$\sim m_f^2/\Lambda^2$.

In both approaches one should try to
construct sensitive
observables to them: as any radiative correction they
contribute to many
observables, but one should identify the appropriate
ones in order
to disentangle them.
 In some observables, for example, like cross sections and
widths, these dipole moments are hidden by tree level
contributions and also mixed up with other quantum
corrections.   One has to deal
with the spin properties in order to obtain sensitive
observables.  The spin density matrix of the produced
lepton/quark pairs has sensitive terms to the dipole
moments, both in the single lepton/quark--polarization and in
the spin--spin correlation terms.  The discrete symmetry
properties of the dipole moments allow a clear search out of
these observables.  For example, in order to disentangle the
WMDM, the parity--odd and time reversal--even single lepton
transverse polarization (within the collision plane)
is found to be a good
candidate \cite{nos,nos1}.
The $CP$--violating WEDM can be found in the single
polarization terms and in correlation terms of the spin
density matrix.  Genuine (\ie the ones that  do not receive
contributions from unitarity corrections and are non--vanishing
only if $CP$--violating pieces appear in the lagrangean)
$CP$--violating observables can be defined from both single
polarization  and from correlations terms, but also non
genuine $CP$--violating observables may be useful in order to
put bounds on these dipole moments.  This paper is organized
as follows:  in section 2 the results appearing in the
literature related to the WMDM are reviewed, in section 3
the WEDM for leptons is studied, and in section 4 it is
shown how the WEDM for quarks, mainly the top quark, can be
bounded.  Section 5 presents some conclusions and the
perspectives on this topic.

\section{Weak--Magnetic Dipole Moment }

 The anomalous WMDM is generated through a chirality flip
mechanism, so it  is expected to be  proportional to the
mass of the particle. Thereby,  only heavy leptons and quark are
good candidates to
have a sizable WMDM: $\bf{\tau}$, ${\bf c}$, ${\bf b}$
 and ${\bf t}$ quarks.

The weak--magnetic dipole moment was investigated for the
$\tau$ in \cite{nos,nos1}, and the WMDM for $b$ is studied in
\cite{nos2}.
The electroweak gauge invariant anomalous WMDM for the $\tau$
was  computed  in \cite{nos1} within the Standard Model,
 to first order in the coupling constant,
 and appropriate observables to
measure it were studied in \cite{nos,nos1}.
There, it is shown  that for $e^+\, e^- \longrightarrow
\tau^+ \tau^-$ unpolarized scattering at the $Z$--peak, the
transverse (within the collision
plane) and normal (to the collision plane)
 single $\tau$ polarizations are  sensitive to the
real and imaginary parts of the
anomalous weak--magnetic  dipole moment,
respectively. Polarization measurements are accessible
for the $\tau$
by means of the energy and angular distribution of
its decay products.

The WMDM is defined in the following way.
The matrix element of the vector neutral current coupled
to the $Z$ is written, using Lorentz covariance, in the form
\beq
\bar{u}(p_-) \,V^{\mu}(p_-,p_+)\, v(p_+)
= e \,\bar{u}(p_-) \left[ \frac{v(q^2) \gamma^\mu}{2 s_w c_w}+i
\frac{a^w_\tau(q^2)}{2 m_\tau}
\sigma^{\mu\eta} q_\eta\right] v(p_+)
\label{mu}\eeq
where $q=p_-+p_+$,  $e$ is the proton charge and
$s_w$, $c_w$
are  the weak mixing  angle sine and cosine,
respectively. The first term $v(q^2)$ is the Dirac
 vertex (or charge radius) form factor
 and it is present at tree level with a value
 $v(q^2)=\frac{1}{2}-2\, s_w^2$, whereas the second form factor
is the WMDM and  only appears
due to quantum corrections.
Only the on--shell vertex with $q^2=M_Z^2$ is entitled to be
electroweak gauge invariant in the Standard Model.

 In order to compute the anomalous WMDM, there are 14 diagrams
to calculate in the t'Hooft--Feynman gauge.
 In Figure 1 a generic diagram is shown;
$\alpha , \beta ,\gamma$  stand for the
particles circulating in the loop: $N\tau^+\tau^-, \nu C^+C^-,
\nu\nu C^-, \tau^-NN^\prime$, where $N,N^\prime
=\gamma , Z, \chi, \Phi$, $N\neq N^\prime$ and
$C=W^\pm ,
\sigma^\pm$ are all the diagrams present in the calculus.
We denote by $\sigma^\pm$ the charged  non--physical Higgs
and by $\chi$ and $\Phi$ the neutral non--physical and
physical ones.
\mxfigura{3cm}{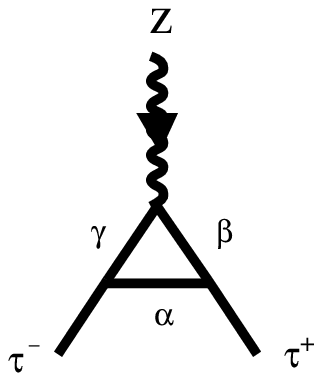}{Contributing Feynman
diagrams to the WMDM
in the t'Hooft--Feynman gauge.}{Figure 1}

There are 6 diagrams that are not present in the analogous
photon vertex
case. these have the following particles circulating in the loop:
 $W\nu\nu$, $\sigma^-\nu\nu$, $\tau Z\Phi$, $\tau \Phi Z$,
$\tau\Phi\chi$, $\tau\chi\Phi$. In fact,
one of these
(the one with $W\nu\nu$ in the loop) gives the leading
 contribution;   this
show that the quantity is governed by  quantum pure--weak
 effects.

All  contributions can be  written as:
\beq
a_{ABC}=\frac{\alpha}{4  \pi} \;\;
\frac{m_\tau^2}{M_Z^2}\;\,\sum_{ij}
c_{ij} {I_{ij}}^{ABC}
\label{a}\eeq
where $A$, $B$ and $C$ are the particles circulating in the
loop, counting
clockwise in the diagrams from the particle between the two
fermion lines,
$c_{ij}$ are
coefficients depending on masses and coupling constants,
and ${I_{ij}}^{ABC} \equiv
I_{ij}(m_\tau^2,q^2,m_\tau^2,m_A^2,m_B^2,m_C^2)$
 are scalar, vector or tensor 3--point functions defined
in  \cite{nos1}.

When computing the diagrams we  only select the
tensor structure related
to the WMDM,
and we  also verify the vector current conservation
 as a check of our expressions
 (there is no induced $(p_-+p_+)^\mu$ term in
Eq.(\ref{mu})).
The external lines are on the mass shell,
\ie $p_-^2=m_\tau^2\,$,
$p_+^2=m_\tau^2$ and $(p_-+p_+)^2=M_Z^2$ respectively.
Some of the diagrams with  the
propagation of Higgs or would--be Goldstone bosons particles
are suppressed by extra ($\frac{m_\tau^2}{M_{Z,\Phi}^2}$) terms
in such a way that the $a_{\chi\tau\tau}, a_{\Phi\tau\tau},
a_{\sigma\nu\nu}$ and $a_{\sigma\sigma\nu}$ contributions to
$a^w_\tau$ are negligible.
 Diagrams in
which the  Higgs and the neutral would--be Goldstone
boson particles
 couple to the $Z$ only contribute to the axial form factor
 and not to the magnetic moment
($a_{\tau\Phi\chi}=a_{\tau\chi\Phi}=0$).

The ${I_{ij}}^{ABC}$ functions were analytically computed
in terms of  dilogarithm functions, and checked with a
numerical
integration for  the $m_\tau \rightarrow 0$ limit.
 Some details are given in \cite{nos1}. They obtain that the
numerical contribution of each diagram is:
\bea
a_{\gamma\tau\tau}=-\frac{\alpha}{4  \pi}
\frac{m_\tau^2}{M_Z^2}
\;(1.32 - 0.52\,  i) \simeq (3.12 - 1.23\, i) \times 10^{-7}
\nonumber\eea
\bea
a\ztautau =\frac{\alpha}{4  \pi} \frac{m_\tau^2}{M_Z^2}
\;(0.17 + 0.08\, i) \simeq (3.92 + 1.88\, i) \times 10^{-8}
\nonumber\eea
\bea
a\nuww =\frac{\alpha}{4  \pi} \frac{m_\tau^2}{M_Z^2}
\;(- 7.07 )
\simeq -1.68 \times 10^{-6}
\nonumber\eea
\bea
a\nuws = \frac{\alpha}{4  \pi} \frac{m_\tau^2}{M_Z^2} \;0.45
\simeq  1.06 \times 10^{-7}
\nonumber\eea
\bea
a\nusw = \frac{\alpha}{4  \pi} \frac{m_\tau^2}{M_Z^2} \;0.45
\simeq  1.06 \times 10^{-7}
\nonumber\eea
\bea
a\tfz = a\tzf =-\frac{\alpha}{4  \pi}
\frac{m_\tau^2}{M_Z^2} \;(0.07\;;\;0.03\;;\;0.02)
\simeq - (0.15\;;\;0.07\;;\;0.04) \times 10^{-7}
\nonumber\label{num}\eea
\beq
 a\wnunu =\frac{\alpha}{4  \pi} \frac{m_\tau^2}{M_Z^2}
\;(- 4.11 - 2.12\, i)
\simeq -\; (0.974 + 0.502\, i) \times 10^{-6}
\eeq
where the  values between parenthesis for $a\tfz = a\tzf$
correspond to $\frac{M_\Phi}{M_Z}=1,2,3$ respectively.

Finally, the value of the computed  WMDM is
\beq
a_\tau^w (M_Z^2)= - \;(2.10 + 0.61\, i) \times 10^{-6}
\label{anom}\eeq
The Higgs mass
only modifies the real part of this result less than a 1\%,
 from
the value $-\,2.12 \times 10^{-6}$ to $-\,2.10 \times
10^{-6}$ for
$1 < \frac{ M_\Phi}{M_Z} < 3$. In Eq.(\ref{anom})  we have
chosen $M_\Phi=2 M_Z$.

We should point out that, contrary to
 the well
known photon--electroweak anomalous magnetic moment,
the non--vanishing absorptive part in Eq.(\ref{anom}) is due to
the fact that we compute on the $Z$ mass
shell $q^2=M_Z^2$, not $q^2=0$.
In fact, one expects a non--vanishing imaginary part coming
from unitarity.

High--precision measurements at LEP/SLC where used in \cite{kopp}
in order to extract bounds for the weak--dipole moments.
They found bounds of the order $10^{-1}-10^{-2}$
for the WMDM of
$\tau, c$ and $b$ quarks, where data from the $Z$--width,
forward--backward
asymmetry and additional angular distributions where used.
However, one should take into account that these
observables are
not the most appropriate ones in order to extract this kind
of information.
 For one hand they receive contributions from any radiative
correction or new
 physics terms, but on the other, they do not depend
linearly on the dipole
 moments.
For example, one should identify observables that
vanish when the
fermion mass (and the dipole moments) vanishes:
these are, in principle, good candidates in order
to measure the dipole moments. For
$e^+e^-\lra \tau^+\tau^-$
 unpolarized collisions, in the $m_f=0$ limit the only
non--vanishing
component of $\tau$--polarization is the longitudinal one.
Then, both the
transverse (within the collision plane) and normal
(to the collision plane)
polarization components vanish in that limit. These ideas
where developped in
\cite{nos,nos1}: when $m_f,a^w_f \ne 0$ then
$P_N^f,P_T^f \ne 0$.
The transverse and  normal single $\tau$ polarization terms
were used to construct asymmetries proportional to
these dipole--moments.
The transverse polarization term in the cross section
is proportional to the real
part of the WMDM, except for a small helicity--flip
suppressed tree level contribution:
\bea
&\displaystyle{\left.\frac{d\sigma}{d\Omega_{\tau^-}}
\right|_T}= &
\frac{\alpha^2\beta}{128\,s_w^3c_w^3\Gamma_Z^2}\;a\;
\sin\theta_{\tau^-}
\bigg\{-\left[2v^2+(v^2+a^2)\beta\cos
\theta_{\tau^-}\right]\frac{v}{\gamma s_wc_w} + \nonumber\\
& &\begin{array}{c} \\ \end{array}
 2\gamma\left[2v^2(2-\beta^2)+(v^2+a^2)\beta\cos
\theta_{\tau^-}\right] Re(a^w_\tau)\bigg\} (s_-+s_+)_x
\eea
The normal polarization term is proportional to the absorptive
part of the WMDM, except for possible WEDM or electric
 dipole moment
contributions:
\bea
\left.\frac{d\sigma}{d\Omega_{\tau^-}}\right|_N= &
\displaystyle{\frac{\alpha^2\beta}{128\,s_w^3c_w^3\Gamma_Z^2}}\;
(-2 v)   \gamma\beta\sin\theta_{\tau^-}\;[2 a^2+(v^2+
a^2)\beta\cos
\theta_{\tau^-}]\;\times\nonumber\\&Im(a^w_\tau)
\; (s_-+s_+)_y
\eea
where $a$ and $v$ are the neutral axial and vector
couplings, $\gamma= \frac{m_Z}{2 m_\tau}$ is the dilation
factor, $\beta$ is the $\tau$ velocity and $\theta_{\tau^-}$
is the angle determined by the $e^-$ and the $\tau^-$
momenta.

The spin properties of the $\tau$ can only be analyzed
from their decay products. In order to have access to
 the single
$\tau$ polarization,   the $\tau$ frame has to be
reconstructed. Micro--vertex detectors allow such
a reconstruction,
 as was shown in \cite{kuhn}, for the case in which
  both $\tau$'s
decay into (at least one) hadrons and their energies
and tracks are
reconstructed.

Let us consider the processes
$e^+e^-\lra\tau^+\tau^-\lra h_1^+Xh_2^-\nu_\tau$ and
$e^+e^-\lra\tau^+\tau^-\lra h_1^+\bar{\nu}_\tau h_2^-X$.
One can construct a sort of mixed
``up--down--forward--backward''
asymmetries in order to disentangle the
dispersive and absorptive
parts of the WMDM.
They select the leading $\cos\theta_\tau \cos\Phi_{h^\mp}$
and  $\cos\theta_\tau \sin\Phi_{h^\mp}$, respectively,
 in  the cross section:
 \begin{equation}
A_{Dis}^\mp=\frac{\sigma^\mp_{Dis}(+)-\sigma^\mp_{Dis}(-)}
{\sigma^\mp_{Dis}(+)+\sigma^\mp_{Dis}(-)}
\end{equation}
with
\begin{eqnarray}
\hspace{-.75cm}\sigma^\mp_{Dis}(\pm)= &\left[\dsp{\int_{0}^{1}
d(\cos \theta_{\tau^-})
\int_{-\pi/2(\pi/2)}^{\pi/2(\frac{3}{2}\pi)} d \phi_{h^\mp}
 + \int_{-1}^{0} d (\cos\theta_{\tau^-})
\int_{\pi/2(-\pi/2)}^{\frac{3}{2} \pi (\pi/2)}d
\phi_{h^\mp}}\right]
\times\nonumber\\ &\dsp{
\frac{d \sigma}{d (\cos\theta_{\tau^-}) \; d \phi_{h^\mp}}}
\end{eqnarray}
and
\begin{equation}
A_{Abs}^\mp=\frac{\sigma^\mp_{Abs}(+)-\sigma^\mp_{Abs}(-)}
{\sigma^\mp_{Abs}(+)+\sigma^\mp_{Abs}(-)}
\label{asc}
\end{equation}
where
\begin{eqnarray}
\hspace{-.75cm}\sigma^\mp_{Abs}(\pm)= &\left[\dsp{\int_{0}^{1}
d (\cos\theta_{\tau^-})
\int_{0(\pi)}^{\pi (2\pi)} d \phi_{h^\mp}
 + \int_{-1}^{0} d (\cos\theta_{\tau^-})
\int_{\pi(0)}^{2 \pi(\pi)}d \phi_{h^\mp}}\right]
\times\nonumber\\ &
\dsp{\frac{d \sigma}{d (\cos\theta_{\tau^-})\; d \phi_{h^\mp}}}
\end{eqnarray}

After some algebra one finds:
\beq
{A_{Dis}}^\mp= \mp \alpha_h \frac{s_wc_w}{4 \beta}
\frac{v^2+a^2}{a^3}\left[
-\frac{v}{\gamma s_wc_w} +2\gamma \; Re(a^w_\tau)\right]
\eeq
\beq
{A_{Abs}}^\mp = \mp \alpha_h \frac{3 \pi\gamma}{4} c_w s_w
\frac{v}{a^2}\;Im(a^w_\tau)
\eeq
where $\alpha_h$ is a parameter that expresses the
sensitivity of
each channel to the $\tau$--spin properties.
The $\mp$ signs refer to the processes defined above.
Collecting events from the $\pi , \rho$ and $a_1$
 channels  it is possible to
put the following bounds:

\begin{equation}
 \abs{Re(a^w_\tau)} \leq 4\cdot 10^{-4}
\label{awt}
\end{equation}
\beq
\abs{Im(a^w_\tau)} \leq 1.1 \times 10^{-3}
\eeq
 These results were obtains considering $10^7\,Z$ events, and
the semileptonic decay channels considered amount to
about 52\% of
the total decay rate.

The Standard Model predictions
 for the real and
imaginary parts of the WMDM are not actually accessible
nowadays. Any signal
coming from  the observables defined above should be
related to new physics.

\section{Weak--Electric Dipole Moment }

In this section the experimental status and the theoretical
results concerning the WEDM for leptons are presented.
Let
us begin by the theoretical expectations.
The time reversal--odd WEDM depends on the underlying physics
of the CP violation mechanisms of the model.
 Standard Model
$CP$--violating effects, as the WEDM for leptons should, in
principle, not be observable within the present experimental
limits: they receive contributions through the
Kobayashi--Maskawa matrix
and only at very high order in the coupling constant.
 This is the main reason to look after them:  a
non--vanishing signal related to them would be a clear claim
for physics beyond the Standard Model.  In many extensions
of the Standard Model and the Kobayashi--Maskawa mechanism,
the $CP$--violation in the lepton sector occurs naturally,
and the generated electric and weak--electric dipole moments
are proportional to the mass of the lepton.  At the $Z$--mass
scale, it is therefore sensible to investigate the heavy
flavours.  A large number of papers are devoted to the study
of the $\tau$ dipole moments, and we will now review their
paramount results.

The subject of $CP$--violation and its related observables has
received much attention in  recent years.
Both ${\bf \tau}$ and ${\bf t}$ have a large branching
ratio in weak decays, so
 $\tau$ and $t$
physics may be studied in a similar way. Theoretical
work done for the tau can be extended so as to be useful
for the top.  Some authors have investigated the
Lorentz structure
of the $\tau -\tau -Z$ vertex, looking for test of
discrete symmetries  and
 possible Standard Model deviations. In
refs.\cite{nelson,blmn,hei}
 some of the results on $\tau$ physics were studied.
More references
can be found in these articles.
The WEDM is defined in the following way.
The matrix element of the axial--vector neutral current coupled
to the $Z$ is written, using Lorentz covariance, in the form
\beq
\bar{u}(p_-) \,A^{\mu}(p_-,p_+)\, v(p_+)
= e \,\bar{u}(p_-) \left[ \frac{a(q^2)
\gamma^\mu\gamma_5}{2 s_w c_w}+
\frac{1}{e}\, d^w_\tau(q^2)\,
\sigma^{\mu\eta}\,\gamma_5\, q_\eta\right] v(p_+)
\label{mue}\eeq
 Again, the first term $a(q^2)$ is the  Dirac
axial vertex  form factor
 and it is present at tree level with a value
 given by the third component of weak--isospin,
 whereas the second form factor
is the WEDM and  only appears
due to quantum corrections at very high orders.
In extended models with scale $\Lambda$, one
expects $d^w_f\sim m_f/\Lambda^2$.

A sizable WEDM would led to a deviation (proportional to the
 square of the WEDM) of the cross section for $e^+e^- \lra
 f\bar{f}$ from its standard value.  In this way Barr and
 Marciano \cite{barr} have taken into account the PETRA
 results for the cross section
  to deduce bounds for the electric dipole moment of
 the tau--lepton.

A WEDM induces \cite{blmn} an additional
 contribution to the $Z$ partial width:
  \beq
 \Delta\Gamma_\tau = \abs{d^w_\tau}^2
 \frac{m_Z^3}{24 \pi}
 \eeq
 From the comparison of the value measured \cite{lep} at LEP
 of the $Z$ partial width $\Gamma_\tau=(84.26\pm 0.34)\;MeV$
 and the Standard Model theoretical prediction \cite{sm}
 $\Gamma_\tau^{SM}=(83.7\pm 0.4)\;MeV$  an upper limit
 \beq
\abs{d^w_\tau}\leq 2.3\times 10^{-17} e \, cm
\eeq
 at
95\% C.L. is obtained.

This argument is an indirect one, and other $CP$--even and
$CP$--odd effects coming from other terms, for example new
physics, may compete.  Moreover, this is certainly not the
most efficient way to put bound on these dipole moments; for
instance, linear effects in the dipole moment (through
 CP--odd observables) allow to put
stringent limits.  One should first understand where the
most important effects coming from these dipole moments
manifest themselves and then, one should either look for
genuine $CP$--violating observables or to look for
observables where effects coming from non $CP$--violating
pieces of the lagrangean are suppressed.

At the $Z$--peak  the electric and weak--electric dipole moments
can be separated in the observables: the last one should be
enhanced at his natural scale, while the first one is suppressed.

All the information one can extract from the process $e^+e^-
\lra \tau^-\tau^+$ is contained in the spin density
matrix.  It has terms that allows to define genuine and
non--genuine $CP$--violating observables.  In \cite{hei} the
first alternative was chosen, whereas in \cite{nos} the
second one was chosen.
\mxfigura{12cm}{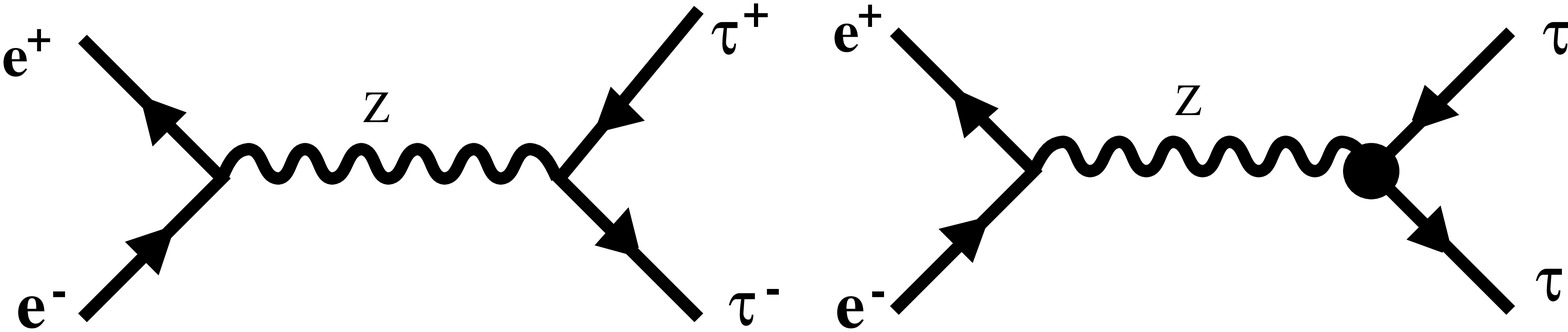}{ Leading Feynman
diagrams considered for processes related to
the WEDM: the left is the tree level diagram while
the right one
is the leading one in the $CP$--violating effective
vertex.}{Figure 2}

In \cite{hei} triple correlation products of momenta were
used to define genuine $CP$--violating observables in
$\tau^\pm$ pair production.  The $\tau$ lepton decays before
reaching the detectors, and only momenta of the $\tau$ decay
products are, in principle, accessible.  This triple
correlation products are generated by the spin--spin
correlation terms in the spin density matrix.  For example,
for the case where both $\tau$'s decay into $\pi\nu$, the
knowledge of the $\hat{q}_\pm=\frac{ q_{\pm}}{\abs{q_\pm}}$
momenta of both $\pi^\pm$ allow to compute the centre of
mass expectation value of the following tensor observables
\bea
T_{ij}=(q_+-q_-)_i (q_+\times q_-)_j+(i\leftrightarrow
j)\nonumber\\ \hat{T}_{ij}=(\hat{q_+}-\hat{q_-})_i \displaystyle{
\frac{(\hat{q_+}\times \hat{q_-})_j} {\abs{\hat{q_+}\times
\hat{q_-}}}}+(i\leftrightarrow j)
\eea
The expectation
value is proportional to the WEDM:
\beq
 <T_{ij}>_{AB}
\simeq \frac{m_Z}{e}\; c_{AB} \; s_{ij}\; d^w_\tau
\label{t}\eeq
 \beq
<\hat{T}_{ij}>_{AB} \simeq \frac{m_Z}{e} \; \hat{c}_{AB} \;
s_{ij} \; d^w_\tau
\label{that}\eeq
where A and B are the decay modes of
$\tau^-$ and $\tau^+$, respectively.  The constants $c_{AB}$
and $\hat{c}_{AB}$ describe the sensitivity of the different
decay channels to the spin properties of the $\tau$, and
$s_{ij}$ is the tensor polarization of the $Z$.  One can
 construct $CP$--odd tensor observables and that are time
reversal even or odd.  In this way the absorptive and
dispersive parts of the WEDM can be tested.  Taking into
account only the Standard Model amplitudes and the first
order WEDM effective vertex (shown in Figure 2)
these tensor observables yield
the following sensitivities.  At the $Z$--peak, and with
$10^7 \;Z$ events an upper limit $1.3 \times 10^{-18} \,e\,
cm$ for the
dispersive part of the WEDM is claimed, whereas for the
absorptive one the limit is $3.2 \times 10^{-17}\,e \,cm$.

These ideas where followed by the OPAL \cite{opal} and ALEPH
\cite{aleph} Collaborations to put bounds on the dispersive
and absorptive parts of the WEDM.  Collecting 20.000
$\tau^\pm$ events from 1990--1992 data, and studying the
decay channels $e\nu\nu, \mu\nu\nu, \pi\nu, \rho\nu,
a_1\nu$, ALEPH found, at the 95\% C.L., the upper bound
\beq
\abs{d^w_\tau}\leq 1.5\times 10^{-17}\,e \, cm
\eeq
Data from
1991-1993 were used by OPAL, altogether resulting in 28000
$\tau^\pm$ pairs to find:
\beq
\abs{Re(d^w_\tau)}\leq 7.8\times
10^{-18}\,e \, cm
\eeq
\beq
\abs{Im(d^w_\tau)}\leq 4.5\times
10^{-17}\,e \, cm
\eeq
 at 95\% C.L., where the decay
channels $l l^\prime, l m$ and $m m\prime$ where $l,l^\prime=e
\mu$ and $m,m^\prime=\pi,\rho, a_1$ where taken
into account (except
for the $a_1 a_1$ decay channel) in order to obtain the
first result, and the channels $ee, \mu\mu, \pi\pi, \pi\rho,
\rho\rho$ were used in the second one.

Similar tensor observables may also be useful when the
 initial state spin density matrix is not $CP$--even.  They
 cease to be genuine $CP$--violating observables.  In
 \cite{indu} the $e^-$ beam was considered with longitudinal
 polarization, and they argue that $CP$--even (suppressed by
 the electron mass) and $CP$--odd (suppressed by a factor
 $\alpha^2 \Gamma_Z^2/m_Z^2$) effects coming from the
 initial state can be discarded.
 In this case the CP--odd, P--odd correlations (\ref{t}) and
 (\ref{that}) are not necessarily proportional to the small
 parity violating parameter $r=2 v^ea^e/((v^e)^2+(a^e)^2)$
 in the electron vertex, and $r$ is replaced by the much higher
 longitudinal polarization $P^e_L$ of the beam, about 70\% at SLC.
  A sensitivity of
  $10^{-17}\,e\,cm$
 would be achieved when $\pi\nu$ and $\rho\nu$ decay
 channels for $10^6\;Z$ events with $e^-$ polarization 62-75
 \% (likely to be available at the SLC at Stanford) is
 supposed.

In \cite{nos} the single $\tau$ polarization pieces of the
spin density matrix were used in order to define observables
sensitive to the WEDM of the tau--lepton.  The normal
polarization of a single $\tau$ is parity--even and time
reversal--odd, and although it is not a genuine
$CP$--violating quantity it enjoys the following virtues:

i)
it gets a contribution from $CP$--conserving interactions
only through the combined effect of both an helicity--flip
transition and the presence of absorptive parts (unitarity
corrections), which are both suppressed in the Standard
Model,

 ii) with a $CP$--violating interaction such as a WEDM,
it gets a non--vanishing value without the need of absorptive
parts,

 iii) as the leading observable effect comes from the
interference of the $CP$--violating amplitude with the
standard amplitude and the observable is P--even, the
sensitivity of the normal polarization to linear terms in
the (dispersive) WEDM is enhanced by the leptonic axial
neutral current Standard coupling and no need of the
suppressed vector coupling of the $Z$ to $\tau$ appears.

One has still the possibility to compare the normal
polarization for $\tau^+$ and $\tau^-$, thus obtaining a
true $CP$--violating observable but with the half of
statistics.  The single $\tau$ normal polarization for the
process is proportional to:
\beq
2a\gamma\beta\sin\theta_{\tau^-} \left[
2v^2+(v^2+a^2)\beta\cos\theta_{\tau^-}\right]
Re(d^w_\tau)
 \label{pn}\eeq
 In order to extract $CP$--violating information
from the normal polarization it is necessary to reconstruct
the $\tau$ direction.  This possibility was studied in
\cite{kuhn}, and (\ref{pn}) appears in
 the $\sin\theta_\tau\cos\theta_\tau\sin\Phi_h$
 distribution of the decay products.
  It is then possible to construct asymmetries
sensitive to the WEDM, as:
\begin{equation}
A^\mp=\frac{\sigma^\mp(+)-\sigma^\mp_(-)}
{\sigma^\mp(+)+\sigma^\mp(-)}
\label{sc}
\end{equation}
where
\begin{eqnarray}
\hspace{-.75cm}\sigma^\mp(\pm)= &\left[\dsp{\int_{0}^{1}
d (\cos\theta_{\tau^-})
\int_{0(\pi)}^{\pi (2\pi)} d \phi_{h^\mp}
 + \int_{-1}^{0} d (\cos\theta_{\tau^-})
\int_{\pi(0)}^{2 \pi(\pi)}d \phi_{h^\mp}}\right]
\times\nonumber\\ &
\dsp{\frac{d \sigma}{d (\cos\theta_{\tau^-})\; d \phi_{h^\mp}}}
\end{eqnarray}
 One gets, in the presence of a non-vanishing $d^w_\tau$,
 for both $\tau^\pm$:
  \beq
   A^\mp= \alpha_h
\frac{\gamma}{2} s_w c_w\frac{v^2+a^2}{a^3}
Re(d^w_\tau)
 \eeq
\beq
A^{CP}\equiv\frac{1}{2}(A^-+A^+)
\eeq
 Although $A^\mp\neq 0$ is not a genuine
 CP--odd term, $A^{CP}\neq 0$ does.
 The factor
$\alpha_h$ describe the sensitivity of the decay channel
with the hadron $h$ to the spin properties of the $\tau$.
The $A^\pm$ asymmetries may receive contribution from
helicity--flipping transitions coming from unitary corrections,
as it was anticipated in i).  These contributions are
negligible and not show in the above expression.  In fact, the
absorptive part of the WMDM is one of them.
The upper
indexes $\pm$ denote that we are collecting events for the
$\mp$ $\tau$ decay channel and that the asymmetry is
constructed using the $\tau^\pm$ decay product.  What is
tested in the second asymmetry is whether the normal
polarization of both taus are different, and this is
certainly a genuine test of $CP$--violation.  Following these
ideas, a sensitivity
\beq
\abs{d^w_\tau} \le 2.3 \times 10^{-18}\; e \, cm
\eeq
 to the
WEDM is found.

\section{Dipole Moments for Quarks}

The evidence for the top quark existence has generated much
excitation and a large amount of theoretical work is
nowadays devoted to it.  All kinds of possible tests of its
properties, in particular possible extensions to new physics
are being currently investigated.  $CP$--violation related to
$t\bar{t}$ production has received much attention.  In this
section we will review some of these results, but mainly
connected to the dipole moments of the top. Standard Model
predictions and the search for possible new physics effects
in top quark production and decay look promising.

The experiments at FNAL \cite{fnal} and precision data from
LEP \cite{lep} are compatible and give evidence for a
heavy top quark with mass around
$m_t\simeq 170-180\, GeV$.  Such a
heavy top has, to a good approximation, the property
\cite{had} that on average, it decays before it can form
hadronic bound states.  The Standard Model prediction is
that the decay $t \lra W\,b$ is predominant for a heavy top.
Some information about its polarization and spin correlation
may be preserved in its decay products.  The spin effects
can be analyzed through the angular correlation of the weak
decay products.  It is then possible to follow similar
approaches as the ones followed in investigating the $\tau$
dipole moments.
However, at the high energies required for
 the $t\bar{t}$ production,
 both the $\gamma$ and $Z$ electric and weak--electric dipole
moments come into the game, and one has to define observables
to disentangle each other.

A heavy top allows that the $CP$ conjugates modes
$t_L \bar{t}_L$ and $t_R \bar{t}_R$ are produced with a big
percentage, contrary to low mass fermions.
  In \cite{pes,phil} an asymmetry sensitive to CP--violation
constructed with
the event rate difference of these modes was considered.
This asymmetry can be measured through the energy spectra of
prompt leptons coming from the decay channel $t \lra W^+ b
\lra l^+\nu b$ and the conjugate one.  The $W^+$ is
predominantly longitudinal, and assuming $V-A$ weak
interaction, the $b$ quark is preferably left--handed. As the
longitudinal $W^+$ is collinear with the top polarization, so
it is the $l^+$ anti--lepton. Above the $t\bar{t}$ threshold
the top is produced with non  zero momentum. As a result of
the Lorentz boost a $l^+$ coming from a $t_R$ has a higher
 energy than the one produced in a $t_L$ decay.
The same happens in the conjugate channel and finally
in the decay of the
pair $t_L\bar{t}_L$ the lepton from $\bar{t}_L$ has a
higher energy
than the antilepton from $t_L$, while in the decay of
$t_R\bar{t}_R$
 the anti--lepton has a higher energy. Therefore the
 energy asymmetry in the
 lepton is sensitive to the asymmetry
 \beq
 A=\frac{N(t_L\bar{t}_L)-N(t_R\bar{t}_R)}{N(t_L\bar{t}_L)+
 N(t_R\bar{t}_R)}\eeq
This asymmetry is $CP\hat{T}$--odd and sensitive to
 the absorptive
part of the
electric and weak--electric dipole moments.

To illustrate the size of  possible CP--violating effects,
 a Weinberg model where the
Higgs's matrix mixes $CP$--even and $CP$--odd scalars
 was considered:
\beq
\cal{L}_{Yukawa} \,=\, -\frac{m_t}{v}\, \overline{t}\,
 ( a L + a^* R ) \,t
\eeq
For a reasonable combination of the mixing parameters,
they estimated that for
the Next Linear Collider at $\sqrt{s}=500\,GeV\;
m_t\simeq 150 \,GeV$ and
$m_H\simeq 100\, GeV$, the lepton asymmetry defined
 above is of the order
$10^{-3}$. This value is accessible with $10^7$
 $t\overline{t}$ pairs!

The up--down asymmetry in the azimuthal angular distribution,
constructed from the  rate difference between the events with
 $l^\pm$
above and below the reaction plane is also a genuine
$CP$--violating signal. This asymmetry is
$CP\widehat{T}$--even and thus
sensitive to
the dispersive part of the electric and weak--electric dipole
moments.

Similar ideas as the ones developed in \cite{hei} for
the $\tau$
lepton where also  apply \cite{heit}  to $t\bar{t}$ production.
Observables constructed from the momenta of the charged leptons
and/or $b$ jets
originated from $t$ and $\bar{t}$ decay may be measurable in
future experiments. These $CP$--odd and $CP\widehat{T}$--even
observables result from
the interference
terms of the $CP$--even amplitudes with $CP$--odd ones, and
are proportional to the dispersive part of the electric and
weak--electric  dipole moments.
The absorptive parts can only contribute to the next order
in the coupling constant, through the interference with absorptive
parts
of one--loop amplitudes. They also argue that
possible $CP$--violating
effects in $t$ and $\bar{t}$ decay do not contribute to leading
order
in perturbation theory. The one standard deviation accuracies
 obtainable  for the
dispersive and absorptive part
of both electric and weak--electric dipole moments
for the top decay channels $t \lra bX_{had}$ and
$t \lra b l^+\nu_l$
is found to be close to $10^{-17} e \, cm$, assuming 10.000
 $t\bar{t}$
 events with $m_t=150 \,GeV$ at $\sqrt{s}=500\, GeV$.

$CP$--violating asymmetries were also studied \cite{indi}
for the process
 $e^+e^-\lra t\bar{t}$ with longitudinally polarized electrons.
The work of \cite{phil} was extended in order to include
polarization
and to disentangle dispersive and absorptive parts of the
electric and weak--electric dipole moments.
With $\sqrt{s}=500 GeV$, an integrated luminosity of
$10 fbn^{-1}$ and polarized electron beams with
$\pm 50 \%$, $90 \%$ C.L. sensitivities of the order
 $10^{-16}-10^{-17} e\,cm$ are obtained.
This is not enough to test many extended
models \cite{pes,phil}
that predicts  dipole moments two or three orders of
magnitude smaller.

Normal polarization
to the production plane in $e^+e^-\lra t\bar{t}$
 was studied in \cite{kan}. In particular the one loop
 QCD correction was
 considered and the effect is induced from unitarity corrections;
  the electroweak contribution is less than the QCD one for
a heavy top at $\sqrt{s}=500 \,GeV$. In this reference
it is also studied
$CP$--violation in the top decay. Different normal polarization for
$t$ and $\bar{t}$  is generated with a $CP$--violating lagrangean.
In particular, the correlation of the azimuthal angles of the
$W^+$ and $W^-$ is sensitive to this $CP$-violation.

\section{Conclusions}

We have discussed some topics related to the dipole
moments of fermions.
Their chirality--flip vertices may provide some insight
into the origin
 of mass. Their property of being dimension--5 operators in the
 effective lagrangean suggest that (in conventional units)
 the anomalous
 magnetic weak--moment and the weak--electric moment are given by
 $a^w_f\sim m_f^2/\Lambda^2$ and $d^w_f\sim e  m_f/\Lambda^2$
 respectively, where $\Lambda$ is the scale of the
 physics involved.

 Within the Standard Model, the weak--magnetic moment for the tau
 $a_\tau^w (M_Z^2)= - \;(2.10 + 0.61\, i) \times 10^{-6}$
 receives its
 dominant contribution from the triangle loops with $\nu WW$
 and $W\nu\nu$.

 There are observables in the process $e^+e^- \lra \tau^+\tau^-$
 which are linear in the weak dipole moments. In
 particular, the transverse
 (within the collision plane) and normal (to the collision plane)
  polarizations
 of single taus contain the information on $a^w_\tau$
 and $d^w_\tau$, respectively.
 These terms manifest themselves in $\cos\Phi_h$ and
 $\sin\Phi_h$ terms,
 respectively,
 of the angular distribution of the (hadron) decay
 product of the tau.

The weak--electric dipole moment appears in the P--odd, CP--odd
spin correlation of both taus $s_+^\tau s_-^\tau$. This
observable
can be searched for in triple correlations for the momenta of
 the decay products.
At LEP, the OPAL and ALEPH experiments have used this method
to put the bounds $\abs{d^e_\tau}\leq 10^{-17}\,e\,cm$. On
the other hand, the P--even, T--odd
normal polarization of single taus has the virtue to
involve the axial (instead of vector) coupling of the electron.

The electric and weak--electric CP--odd dipole moments for
the $t$--quark
can be searched for by means of CP--odd observables
 at the Next Linear Collider.
\section*{Acknowledgements}

J.B. is indebted to B.Kniehl for the invitation
to the Ringberg Workshop
where we enjoyed a very stimulating atmosphere. G.A.G.S.  thanks
the Generalitat Valenciana for a grant at
 the University of Valencia.  This work has been supported
 in part by CICYT, under Grant AEN 93-0234, and by I.V.E.I..


\begin{thebibliography}{99}

\bibitem{sch} J.Schwinger,\pr{73}{48}{416}.

\bibitem{mar} W.Marciano, Brookhaven National--Laboratory
 preprint, BNL--{61141}, 1994.

\bibitem{nos} J.Bernab\'eu, G.A.Gonz\'alez--Sprinberg and
J.Vidal, \pl{B326}{94}{168}.

\bibitem{nos1} J.Bernab\'eu, G.A.Gonz\'alez--Sprinberg,
 M.Tung and
J.Vidal,
\np{B436}{95}{474}.

\bibitem{kopp} G.K\"opp \etal, DESY Preprint 94-148.


\bibitem{kuhn} J.H.K\"{u}hn, \pl{B313}{93}{458}.


\bibitem{nos2} J.Bernab\'eu, G.A.Gonz\'alez--Sprinberg and
J.Vidal, in preparation.

\bibitem{nelson} C.A.Nelson \pr{D40}{89}{123},
\pr{D41}{90}{2327}(E);\\
C.A.Nelson \pr{D41}{90}{2805};
\\S.Goozovat and
C.A.Nelson \pl{B267}{91}{128};\\S.Goozovat and
C.A.Nelson \pr{D44}{91}{2818}.

\bibitem{blmn} W.Bernreuther \etal , \zp{C43}{89}{117}.

\bibitem{hei} W.Bernreuther and O.Nachtmann,
\prl{63}{89}{2787} and  \prl{64}{90}{1072}(E);\\
W.Bernreuther \etal, \zp{C52}{91}{567};\\ W.Bernreuther,
O.Nachtmann and P.Overmann, \pr{D48}{93}{78}.



\bibitem{barr} S.Barr and W.Marciano in {\em CP Violation
--Advanced Series in H.E.P.  Vol.3}, edited by C.Jarlskog,
World Scientific, Singapore, 1989.



\bibitem{lep} D.Schaile, {\em
Proceedings of the XXVII International Conference on
High Energy Physics},
Glasgow 1994, edited by P.J.Bussey and I.G.Knowles,
 IOPP Publishing Ltd., Bristol and Philadephia, vol.I, p.27 (1995).

\bibitem{sm} W.H\"ollik in {\em Proceedings of the XVI
International Symposium on Lepton and Photon Interactions},
Ithaca 1993, edited by P.Drell and D.Rubin, AIP, New York,
352 (1994).




\bibitem{opal}OPAL Collaboration, R.Akers \etal ,
CERN-PPE/94-171.

\bibitem{aleph} ALEPH Collaboration, D.Buskulic \etal ,
CERN-PPE/944-175.

\bibitem{indu} B.Ananthanarayan and S.D.Rindani,
\prl{73}{94}{1215};\\
\pr{D50}{94}{4447}.

\bibitem{fnal} CDF Collaboration, F.Abe \etal , \prl{73}{94}{225}.


\bibitem{had} I.Bigi and H.Krasermann, \zp{C7}{81}{127};
\\J.K\"uhn, {\it Acta Phys.  Austr.  Suppl.}  {\bf XXIV}
(1982) 203;\\ I.Bigi \etal , \pl{B181}{86}{157}.



\bibitem{pes} C.R.Schmidt and M.E.Peskin,
\prl{69}{92}{410};\\ C.R.Schmidt, \pl{B293}{92}{111}.

\bibitem{phil} D.Chang, W.-Y.Keung and I.Phillips,
 \np{B408}{93}{286}; \np{B429}{94}{255}(E).

\bibitem{heit} W.Bernreuther \etal , \np{B388}{92}{53}.

\bibitem{indi} P.Poulose and S.D.Rindani, PRL-TH-94/31 prepint.

\bibitem{kan} G.L.Kane, G.A.Ladinsky and C.-P.Yuan,
\pr{D45}{92}{124}.

%\bibitem{stie} U.Stiegler, \zp{C57}{93}{511}.


%\bibitem{tsai} Y.S.Tsai, \pr{D4}{71}{2821} and erratum
%\pr{D13}{76}{771}.

%\bibitem{wag} J.H.K\"{u}hn and F.Wagner, \np{B236}{84}{16}.

%\bibitem{rouge} A.Roug\'e, \zp{C48}{90}{75}.

%\bibitem{martin} K.Hagiwara, A.D.Martin and D.Zeppenfeld,
%\pl{B235}{90}{198}.

%\bibitem{fuj} K.Fujikawa, B.W.Lee and A.I.Sanda,
%\pr{D6}{72}{2923}.

%\bibitem{queijeiro} A.Queijeiro, \zp{C60}{93}{667}.


%\bibitem{pass} G.t'Hooft and M.Veltman, \np{B153}{79}{365};
%G.Passarino and M.  Veltman, \np{B160}{79}{151}.

\end{thebibliography}
 \end{document}